\documentclass[a4paper]{article}
\usepackage[british]{babel} % set babel to british rather than american
\usepackage[utf8]{inputenc} % utf8 support in source code
\usepackage[T1]{fontenc} % better support for special characters in pdf but messes up title fonts, can be fixed by installing the debian package cm-super
\usepackage{textcomp}
\usepackage{placeins}

\usepackage[a4paper,top=3cm,bottom=2cm,left=3cm,right=3cm,marginparwidth=1.75cm]{geometry}

\usepackage{authblk} %for nice, normal, author lists. we should be so lucky

\usepackage{subcaption}

\usepackage{xcolor} % for colorized fonts

\usepackage{hyperref}
\usepackage{amsfonts}
\usepackage{amssymb}
\usepackage{booktabs} % fancy tables

\usepackage{afterpage}

\usepackage{siunitx}
\DeclareSIUnit\radlen{\text{\ensuremath{X_{\mathrm{0}}}}}
\DeclareSIUnit\clight{\text{\ensuremath{c}}} % remove 0 subscript from speed of light

\usepackage[labelfont=bf]{caption} %make caption label bold

\usepackage{graphicx}
\usepackage{soul}

\begin{document}

\begin{center}
	
    {\Large \bf First Operation of a Resistive Shell\\}
    {\Large \bf Liquid Argon Time Projection Chamber -\\}
    \vspace*{0.2cm}
    {\Large \bf A new Approach to Electric-Field Shaping}
	\vspace*{1.0cm}
	\setcounter{footnote}{0}  
	\def\A{\kern+.6ex\lower.42ex\hbox{$\scriptstyle \iota$}\kern-1.20ex a}
	\def\E{\kern+.5ex\lower.42ex\hbox{$\scriptstyle \iota$}\kern-1.10ex e}
	\small
	\newcommand{\Aname}[2]{#1}
	\def\titlefoot#1{\vspace{-0.3cm}\begin{center}{\bf #1}\end{center}}
	
	\Aname{Roman Berner\footnote{Corresponding author: \href{mailto:roman.berner@lhep.unibe.ch}{roman.berner@lhep.unibe.ch}}}{Bern},
	\Aname{Yifan~Chen}{Bern},
	\Aname{Antonio~Ereditato}{Bern},
	\Aname{Patrick~P.~Koller}{Bern},
	\Aname{Igor~Kreslo}{Bern},\\
	\Aname{David~Lorca}{Bern},
	\Aname{Thomas~Mettler}{Bern},
	\Aname{Francesco~Piastra}{Bern},
    \Aname{James~R.~Sinclair}{Bern}, and
    \Aname{Michael~S.~Weber}{Bern}
    \vspace*{0.1cm}
	\titlefoot{Albert Einstein Center for Fundamental Physics,\\Laboratory for High Energy Physics, University of Bern, Switzerland\label{Bern}}
	
	\Aname{Ting~Miao}{FNAL}
    \vspace*{0.1cm}
	\titlefoot{Fermi National Accelerator Laboratory, Batavia, IL 60510, USA\label{FNAL}}
\end{center}
\vspace*{1cm}

%%%%%%%%%%%%%%%%%%%%%%%%%%%%%%%%%%%%%%%%%%%%%%%%%%%%%%%%%%%%%
%% ABSTRACT
%%%%%%%%%%%%%%%%%%%%%%%%%%%%%%%%%%%%%%%%%%%%%%%%%%%%%%%%%%%%%
\begin{abstract}
We present a new technology for the shaping of the electric field in Time Projection Chambers (TPCs) using a carbon-loaded polyimide foil.
This technology allows for the minimisation of passive material near the active volume of the TPC and thus is capable to reduce background events originating from radioactive decays or scattering on the material itself.
Furthermore, the high and continuous electric resistivity of the foil limits the power dissipation per unit area and minimizes the risks of damages in the case of an electric field breakdown.
Replacing the conventional field cage
%made of conductive electrodes set at fixed voltage
with a resistive plastic film structure called ``shell'' decreases the number of components within the TPC and therefore reduces the potential points of failure when operating the detector.
A prototype liquid argon (LAr) TPC with such a resistive shell and with a cathode made of the same material was successfully tested for long term operation with electric field values up to about \SI{1.5}{\kilo\volt\per\cm}.
The experiment shows that it is feasible to successfully produce and shape the electric field in liquefied noble-gas detectors with this new technology.
\end{abstract}

\hspace{5cm}

\begin{center}
\end{center}

\newpage
%%%%%%%%%%%%%%%%%%%%%%%%%%%%%%%%%%%%%%%%%%%%%%%%%%%%%%%%%%%%%
%% INTRODUCTION
%%%%%%%%%%%%%%%%%%%%%%%%%%%%%%%%%%%%%%%%%%%%%%%%%%%%%%%%%%%%%
\section{Introduction}
% MOTIVATION FOR NOBLE LIQUID TPCs
% ---------------------------------------------------------------
The high tracking accuracy and the calorimetric features have made Time Projection Chambers (TPCs) based on liquefied noble gases an established detector type for many experiments running or planned on neutrino physics~\cite{SBN_program, DUNE_IDR_Vol1_2018, nEXO_experiment} and direct dark matter searches~\cite{MAX_experiment, XENON1T_experiment, DARWIN_experiment}.
%The high accuracy of tracking and calorimetry have made Time Projection Chambers (TPCs) based on liquefied noble gases a detectro of choice in current neutrino, $0\nu\beta\beta$ and direct dark matter searches~\cite{MicroBooNE_experiment, MAX_experiment, EXO_experiment, XENON1T_experiment, DARWIN_experiment} and future experiments~\cite{DUNE_IDR_Vol1_2018}.
%Modern detectors in neutrino and dark matter experiments are, to a large extent, represented by liquefied noble gas Time Projection Chambers (TPCs).
%This is mainly due to the excellent ionisation and scintillation yields of the detection medium, that allows for high accuracy particle tracking and calorimetry, as well as to the very high breakdown electric field, which allows for electric drift fields as high as some kilo volts per centimetre.
%Time Projection Chambers (TPCs) based on liquefied noble gases are playing a fundamental role in neutrino, $0\nu\beta\beta$ and direct dark matter searches~\cite{MicroBooNE_experiment, MAX_experiment, EXO_experiment, XENON1T_experiment, DARWIN_experiment}.

\begin{figure}[htb!]
    \centering
    \includegraphics[draft=false,width=0.65\textwidth]{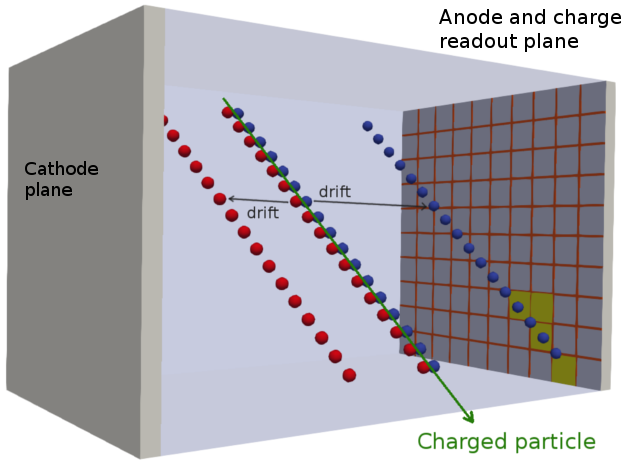}
    \caption{Illustration of the working principle of a TPC. Ionisation electrons (blue) and the ionised atoms (red) drift to the anode and the cathode, respectively.}
    \label{fig: TPC working principle}
\end{figure}

The basic working principle of a TPC is illustrated in Figure~\ref{fig: TPC working principle}:
Charged particles travelling through a medium loose their energy by ionising its atoms.
%as charged particles travel through the medium they loose energy by ionizing the atoms.
Due to the ionisation prompt scintillation light is produced.
An electric field applied between an anode and a cathode placed in the medium (gas or liquid) forces the ionisation electrons to move (drift) towards a charge-collection device at the anode.
Combining charge readout with the time signal from the scintillation light allows for a full 3D reconstruction of the events.
%The basic working principle of a TPC is illustrated in Figure~\ref{fig: TPC working principle}.
%An incident charged particle interacts with the atom(s) of the noble liquid and produces both secondary particles (charged and neutral) and scintillation light copiously.
%The prompt scintillation light is proportional to the energy deposit of the incident particle and is collected by photon detectors to determine the time of the interaction, $t_0$.
%The secondary charged particles travel through the liquefied noble gas and mainly loose energy by ionizing the atoms, knocking out their loosely bound electrons.
%Furthermore, scintillation light is produced.
%An applied electric field between anode and cathode forces the ionisation electrons to move towards the anode.
%Using a charge readout combined with the time signal from the scintillation light allows for a 3D reconstruction of event vertices and particle tracks.
%Furthermore, the track's energy deposition can be used for calorimetry and particle identification.

Conventional TPC designs~\cite{MicroBooNE_experiment,ICARUS_detector,XENON1T_experiment} employ metallic field cage structures to produce a uniform electric field between the anode and the cathode;
%Traditional TPC designs (e.g. used in MicroBooNE~\cite{MicroBooNE_experiment}, ICARUS~\cite{ICARUS_detector}, XENON1T~\cite{XENON1T_experiment}) employ metallic field cage structures to produce a uniform electric field within the active TPC volume.
these consist of a sequence of conductive elements surrounding the TPC drift region.
%These structures consist of a sequence of conductive elements surrounding the TPC drift region between the anode and the cathode.
The segments are electrically connected by resistors that produce the required voltage drop between consecutive field cage units.

%The next generation long baseline neutrino experiments will require detectors with masses $\mathcal{O} \left( 10 \right) \,\mathrm{kt}$, with drift lengths of meters \cite{DUNE_IDR_Vol1_2018}.
The next generation of experiments will require detectors with active volume masses ranging from
$\mathcal{O} \left( 10^3 \right) \,\mathrm{kg}$ for neutrino-less double beta decay studies %\cite{nEXO_experiment}
and direct dark matter searches
%\cite{DARWIN_experiment}
up to $\mathcal{O} \left( 10^7 \right) \,\mathrm{kg}$ for long baseline neutrino experiments, with drift lengths up to several meters.
%\cite{DUNE_IDR_Vol1_2018}.

%With such dimensions cathode bias voltages up to
%\SI{600}{\kilo\volt}
%$\mathcal{O} \left( 600 \right) \,\mathrm{kV}$
%\cite{bib: DUNE TDR Vol.3}
%exceeding
%\SIrange{100}{150}{\kilo\volt}
%are required in order to produce electric fields that can efficiently drift the ionisation electrons over the entire drift length.
%The typical electrostatic energy stored within the TPC is
%$\mathcal{O}\left(100\right) \,\mathrm{J}$
%\cite{DUNE_CDR_Vol4_2016}, hence the occurrence of accidental discharges between the cathode or one of the closest field cage elements and the grounded environment around the field cage might pose serious challenges for the integrity of the electronics used for light and charge detection.
The electrostatic energy stored in a TPC of such dimensions can be up to $\mathcal{O} \left( 100 \right) \,\mathrm{J}$ \cite{DUNE_CDR_Vol4_2016}.
Hence, the occurrence of accidental discharges between the cathode or one of the closest field cage elements and the grounded environment around the field cage poses a serious risk of damage to the readout electronics.
%Future long-baseline neutrino searches (e.g. DUNE) will see LArTPCs deployed on a very large scale of active medium, $\mathcal{O} \left( \text{kt} \right)$, and in high-multiplicity environments, $\mathcal{O} \left( 0.1 \right)$ events per tonne of target medium and per beam spill.
%Kilotonne-scale experiments have drift lengths of $\mathcal{O} \left( \text{m} \right)$ and require very high cathode bias voltages (HV) to maintain a drift field, $\mathcal{O} \left( 10^3~\text{V/cm} \right)$, and with typical stored energies of $\mathcal{O} \left( 100 \text{~J} \right)$~\cite{DUNE_CDR_Vol4_2016}.

The ArgonCube LAr TPC detector concept~\cite{ArgonCube_LOI} was developed among other goals to address the risks associated with high voltage (HV) \cite{ElectricBreakdown_paper} and long drift distances \cite{RnD_at_LHEP, ARGONTUBE}
%carefully studied with the predecessor ArgonTube project \cite{ArgonTube}
through modularisation; separating the detector volume into a number of self-contained TPCs sharing a common cryostat.
Modularisation reduces the stored energy, simplifies electric-field stability and lowers the requirements for LAr purity.
Additionally, it brings the benefit of a reduced track multiplicity per TPC unit which simplifies the event reconstruction.
%Modularisation has the additional benefit of lowering the track multiplicity per TPC, which simplifies the event reconstruction.
This approach requires all components of the TPC to be as compact as possible to maximise the active volume.
Advancements have already been made in the charge and light readout~\cite{first_pixel_paper, LArPix, ArCLight}; here we present a new approach for field shaping.
%The Laboratory of High Energy Physics (LHEP), University of Bern, is involved since 2007 in an extensive research and development (R\&D) program for the development of Liquid Argon TPCs (LArTPC) for neutrino detection, that approached to the ArgonCube project~\cite{ArgonCube_LOI}.
%The Laboratory of High Energy Physics (LHEP), University of Bern, is involved in the ArgonCube research and development (R\&D) program, with the purpose of developing a modular multi-tonne scale LArTPC based detector, suitable for neutrino detection in high interaction rate environments such as in a neutrino beamline near detector.
%The segmentation of large detection volumes with arrays of small independent TPCs offers the advantage of reducing the required HV and the challenges related to it, providing a uniformly instrumented detection volume.
%In comparison to traditional designs, this solution allows for a lower track multiplicity per TPC.
%In addition, the shorter electron drift times allow for an improved accuracy of track reconstruction and calorimetry.

We propose as an alternative to the conventional field cage a continuous resistive plane forming a so-called ``resistive shell''.
This will provide a continuous linear potential distribution along the drift direction, paired with a rather simple mechanical design.
%mechanics
By eliminating the resistor chain the component count is drastically reduced and therefore also the number of potential points of failure.
In the case of an electric breakdown a resistive shell will limit the power release.
The continuous voltage drop from the anode to the cathode requires no electric-field optimisation of the passive electric components as in a conventional field cage.
%The replacement of the resistor chain and of the field-shaping segments with a resistive shell reduces the TPC complexity and the potential points of failure.
%Additionally, the conduction properties of the resistive foil strongly mitigates the amount of charge that could be released in an undesired electric discharge.
%, that would be promptly dumped during its rising phase. 
%Because of this, the challenges posed to the very sensible electronics for the charge and light readout are strongly reduced.
%Finally, thanks to the continuous voltage drop from the anode to the cathode, no field based optimisation of the passive electric components as for traditional field cages is required.
%For a material to be suitable for use as a resistive shell it must have a uniform sheet resistance of $\mathcal{O} \left( 1 \right) \,\mathrm{G}\Omega\mathrm{sq}^{-1}$ at the desired field intensity and temperature of \SI{1}{\kilo\volt\per\cm} and \SI{87}{\kelvin}.
At a temperature of \SI{87}{\kelvin} a material must have a uniform sheet resistance of $\mathcal{O} \left( 1 \right) \,\mathrm{G}\Omega\mathrm{sq}^{-1}$ to be suitable for use as a resistive shell at the desired field intensity of \SI{1}{\kilo\volt\per\cm}. 
The resistive shell can be produced with solely resistive material in the form of a $\mathcal{O} \left( 100 \right)\,\mu\mathrm{m}$ thick foil or as a laminate with a $\mathcal{O} \left( 10 \right) \,\mu\mathrm{m}$ resistive layer on a dielectric substrate.
%In this work we propose
%To address these challenges we propose
%a novel TPC design, where the cathode and the field-shaping cage are replaced by a thin shell consisting of a highly resistive foil with a resistance of
%$\mathcal{O}\left(10^9\right) \,\mathrm{\Omega\text{/square}}$
%at a LAr temperature.
%The high resistivity featured by this material is fundamental to achieve the low-power dissipation at very high cathode bias voltages required to operate a LArTPC in stable conditions.
%This is mainly due to the uniform power dissipation over the whole resistive shell, with no points where local power dissipation (e.g. resistors), as for field cages, might induce more easily boiling of the LAr.
%Such an occurrence would locally increase the electron diffusion coefficient potentially spoiling the position reconstruction accuracy for those electrons with drifting paths close to the ‚hot‘ point.

The TPC presented here represents a proof of principle.
%and a first step toward a future design envisaged for ArgonCube~\cite{ArgonCube_LOI}.
%, its implementation is not indicative of the design envisage for ArgonCube.
The experimental setup of the
%resistive shell
prototype TPC used for the measurements is detailed in Section~\ref{sec: Setup}.
The results of a data taking period of \SI{5}{\mathrm{days}} carried out in July 2018 are presented in Section~\ref{sec: Results}, followed by Section~\ref{sec: Conclusion} with the conclusion and outlook.
\section{Setup of the prototype TPC}
\label{sec: Setup}
\begin{figure}[htb!]
    \centering
    \includegraphics[draft=false,width=0.85\textwidth]{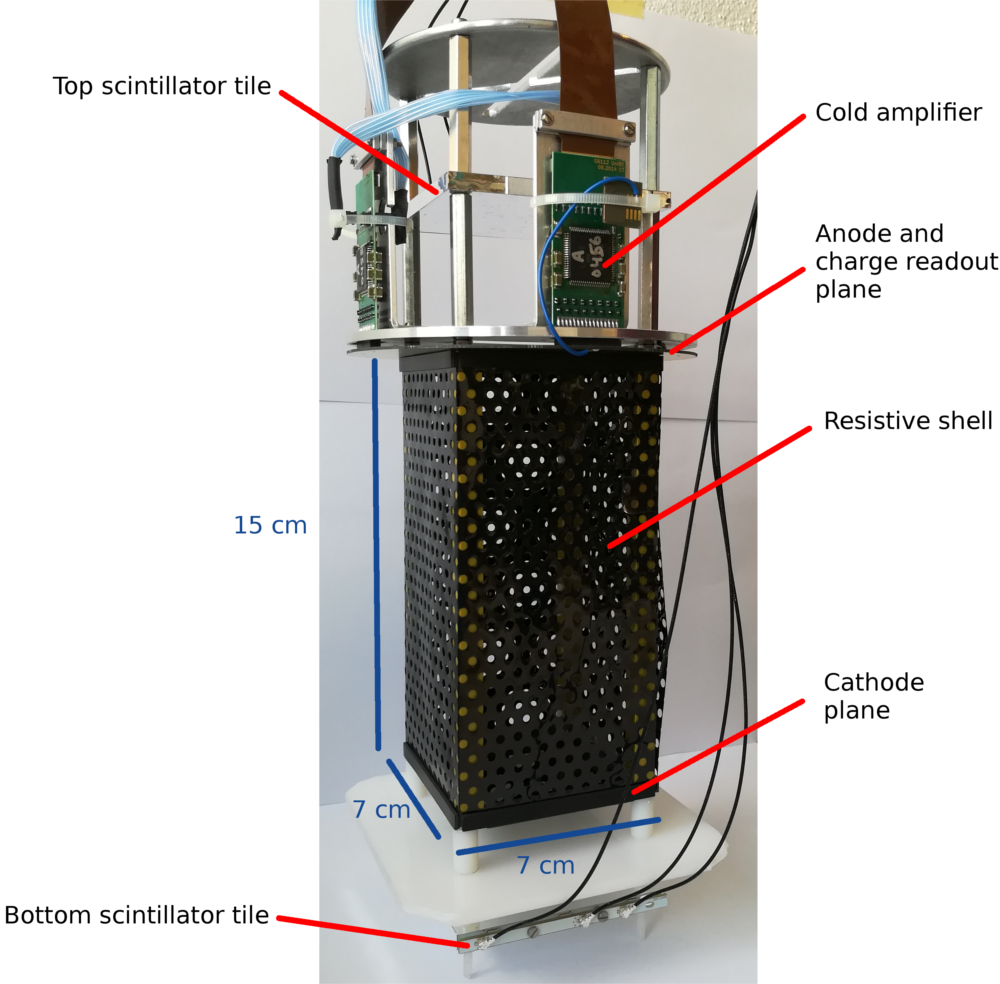}
    \caption{Prototype of the resistive shell TPC used for the measurements. The holes in the shell foil assure an adequate flow of purified LAr from outside into the TPC drift volume.
%Two scintillator tiles, each populated with three silicon photomultipliers, are used to trigger the charge readout system.
%The holes in the resistive shell allow for an appropriate LAr circulation to maintain an uniform LAr purity.
    }
    \label{fig: RSTPC setup}
\end{figure}

The resistive shell TPC prototype built at LHEP\footnote{Laboratory for High Energy Physics}, University of Bern, is shown in Figure~\ref{fig: RSTPC setup}.
The TPC has a \SI{7x7}{\centi\metre} footprint and a \SI{15}{\centi\metre} drift length with the resistive shell constructed from a single sheet of $\sim\,$\SI{50}{\micro\metre} thick carbon-loaded Kapton\footnote{DuPont\texttrademark, Kapton\textregistered \, polyimide film, E. I. du Pont de Nemours and Company, \url{www.dupont.com}.} foil.
The cathode plane is made of the same material.
%The resistive shell TPC prototype built at LHEP
%\footnote{Laboratory for High Energy Physics, University of Bern}, shown in Figure~\ref{fig: RSTPC setup}
%with main components labelled
%, has a
%\SI[product-units=repeat]{15x7x7}{\cm}
%$15 \times 7 \times 7$~cm$^3$
%active volume surrounded by the resistive shell.
%Both cathode and the resistive shell consist of a
%\SI{50}{\micro\metre}
%$50 \text{~$\mu$m}$
%thick foil of Kapton\footnote{DuPont\texttrademark, Kapton\textregistered \, polyimide film, E.~I. du Pont de Nemours and Company, \url{www.dupont.com}.}.
%This foil was punched with a regular pattern in order to maintain an high and homogeneous LAr purity within the TPCs active volume.
The shell as well as the cathode were manufactured in the US before their deployment in Bern.
In this application the resistive shell was perforated to allow for an adequate circulation of purified LAr in the TPC volume.
%The foil of the resistive shell was punched with a regular pattern in order to ease the LAr circulation throughout the TPC boundaries, providing an efficient removal of electronegative impurities from the active volume.

\begin{figure}[htb!]
    \centering
    \begin{subfigure}{0.489\textwidth}
        \includegraphics[draft=false,width=\textwidth]{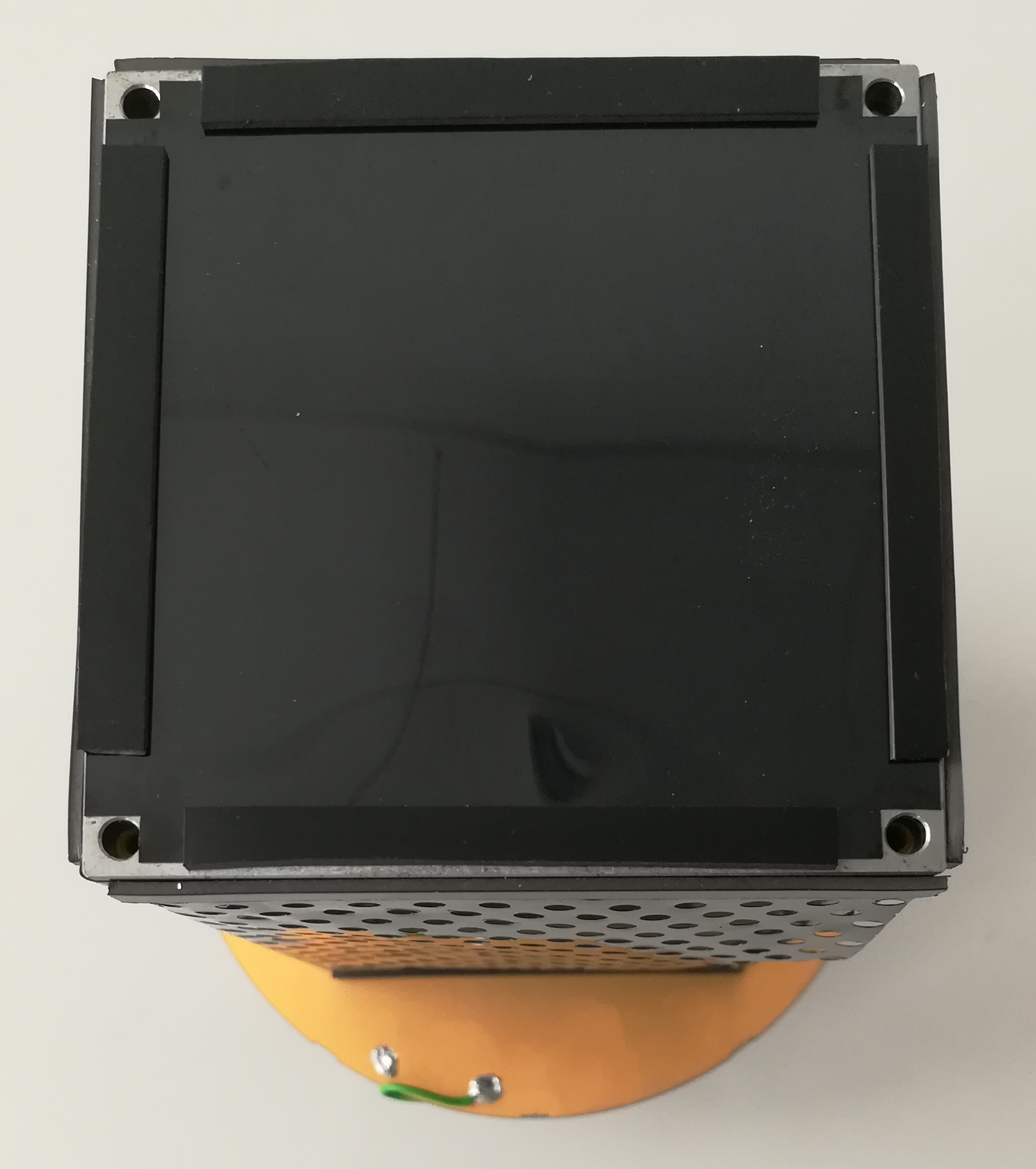}
        %\caption{Mockup 1}
        %\label{fig: Mockup 1}
    \end{subfigure}
    ~ %add desired spacing between images, e. g. ~, \quad, \qquad, \hfill etc. (or a blank line to force the subfigure onto a new line)
    \begin{subfigure}{0.489\textwidth}
        \includegraphics[draft=false,width=\textwidth]{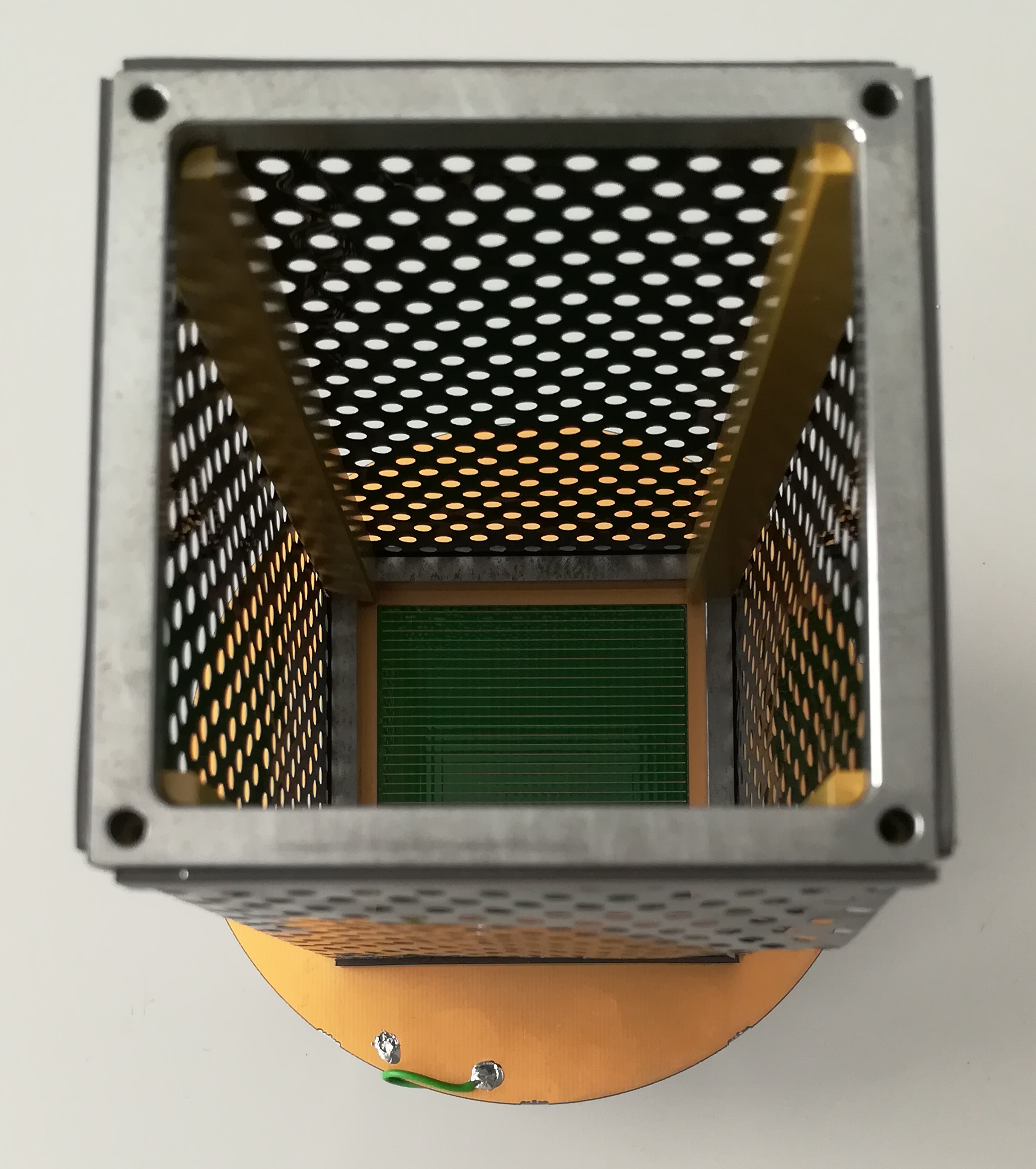}
        %\caption{Mockup 2}
        %\label{fig: Mockup 2}
    \end{subfigure}
    \caption{Resistive shell TPC mockup with and without the attached cathode plane. The resistive foil is mounted with magnet stripes to the grey frames, both of them \SI{5}{\mm} thick.}
    \label{fig: RSTPC mockup}
\end{figure}

The structure of the TPC is provided by a frame as shown in Figure~\ref{fig: RSTPC mockup}.
The frame consists of mild steel squares at both the anode and the cathode with the squares separated by four sections of Polyamide-imide that run through the length of the TPC.
The resistive shell is attached with permanent magnet stripes to a top and a bottom steel square which provide the electrical connection of the shell to the anode and the cathode. 
Incorporating a conducting steel frame reduces the actual length of the resistive shell to \SI{13.5}{\centi\metre}.
Dividing this length by the shell's \SI{28}{\centi\metre} periphery gives a square of $0.482$~sq.

\begin{figure}[htb!]
    \centering
    \begin{subfigure}{0.334\textwidth}
        \includegraphics[draft=false,width=\textwidth]{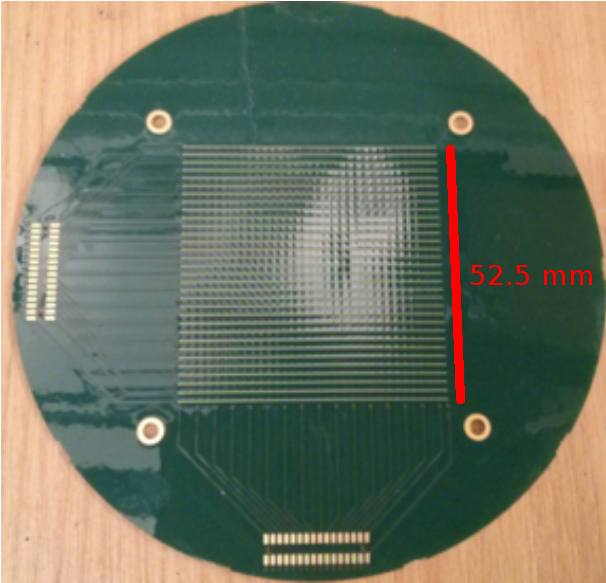}
        %\caption{Top scintillator}
        \label{fig: WirePlane}
        %\caption{}
    \end{subfigure}
    ~ %add desired spacing between images, e. g. ~, \quad, \qquad, \hfill etc. 
      %(or a blank line to force the subfigure onto a new line)
    \begin{subfigure}{0.6452\textwidth}
        \includegraphics[draft=false,width=\textwidth]{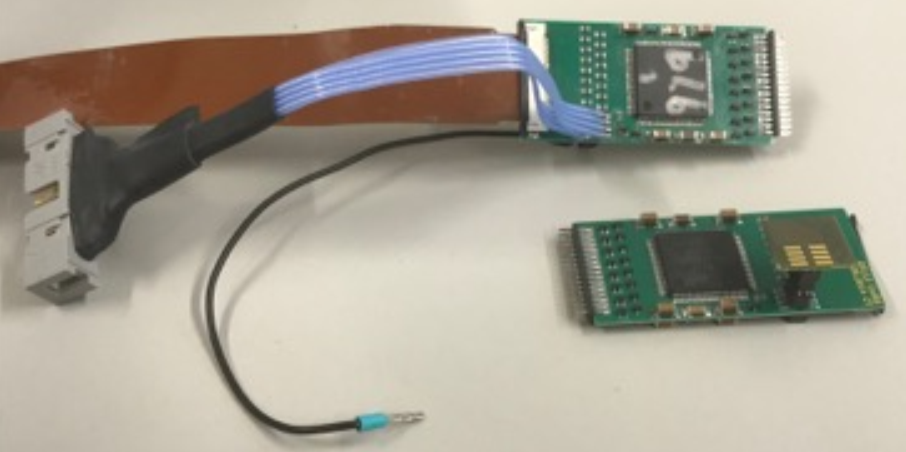}
        %\caption{Bottom scintillator}
        \label{fig: LARASIC}
        %\caption{}
    \end{subfigure}
    \caption{Charge-readout system consisting of $64$~conducting strips printed on two planes, separated by
    \SI{50}{\micro\metre}
    of Kapton and four LARASICs mounted on two PCBs for the charge readout.}
    \label{fig: ChargeReadout system}
\end{figure}

The TPC charge-readout system consists of two planes of conducting strips printed as parallel lines on either side of a \SI{50}{\micro\metre} thick Kapton foil, as shown in Figure~\ref{fig: ChargeReadout system}.
Each of those planes, collection and induction, are oriented at \SI{90}{\degree} and count $32$ paths with a pitch of \SI{1.7}{\mm}.
%Each plane has $32$ wires at a pitch of \SI{1.7}{\mm}, and is printed as traces on either sides of a \SI{50}{\micro\metre} thick Kapton foil, as shown in figure~\ref{fig: ChargeReadout system}.
%The planes are printed on the two sides of a
%\SI{50}{\micro\metre}
%thick Kapton foil, as shown in figure~\ref{fig: ChargeReadout system}, and are coupled to cryogenic charge preamplifiers: one LARASIC4* and three LARASIC7 \cite{LARASIC}, each providing the simultaneous amplification of the charge signals for $16$~different wires.
%Since the two planes are separated by the dielectric Kapton, the charge readout configuration of the prototype TPC is opposite to that usually employed in LArTPCs, where the charge is measured with two or more wire planes.
%Namely, the collection wires facing the sensitive volume, where the electrons are neutralised at their arrival, and the induction wires printed on the other side of the Kapton.
Both planes are coupled to four cryogenic preamplifiers in total: one LARASIC4* \cite{LARASIC} and three LARASIC7 each providing the amplification for $16$ channels.
%Two custom made printed circuit boards, each hosting two LARASIC chips, provide the circuitry for the connection of the amplifier input channels to the two wire planes.
%The amplified signals are then transmitted from the boards by two flat cables, each hosting $32$~channels, to a custom feedthrough where the $64$~signals are doubled into a positive and negative copy for their differential transmission outside of the cryostat.
The data acquisition (DAQ) was altered from that described in Ref.~\cite{protoLASER} in order to include differential signalling.
Ground loops are avoided by not connecting the signal sink to the same ground as the signal source.
%because the signal sink does not need connecting to the same ground as the signal source.

\begin{figure}[htb!]
    \centering
    \begin{subfigure}{0.469\textwidth}
        \includegraphics[draft=false,width=\textwidth]{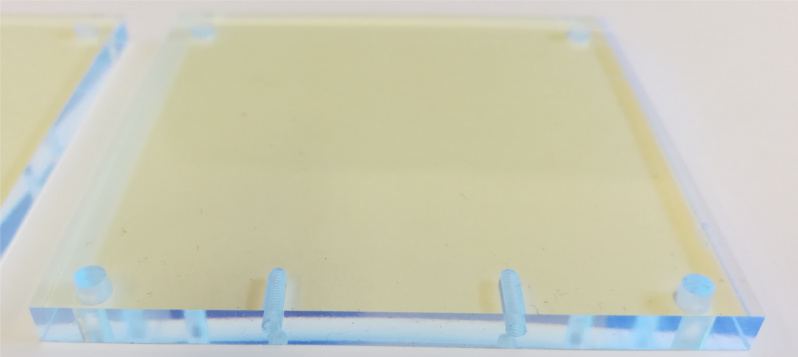}
        %\caption{Top scintillator}
        \label{fig: Scintillator bare}
        %\caption{}
    \end{subfigure}
    ~ %add desired spacing between images, e. g. ~, \quad, \qquad, \hfill etc. 
      %(or a blank line to force the subfigure onto a new line)
    \begin{subfigure}{0.51\textwidth}
        \includegraphics[draft=false,width=\textwidth]{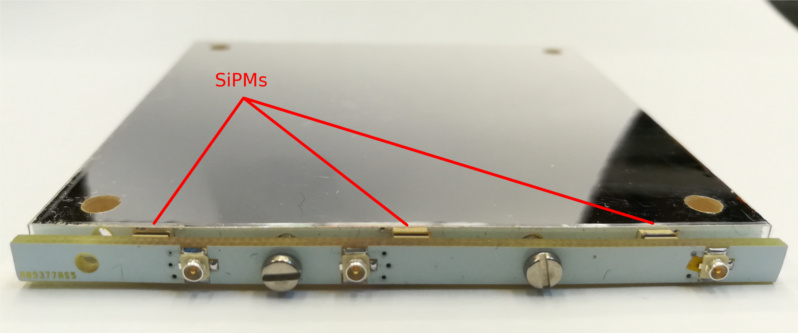}
        %\caption{Bottom scintillator}
        \label{fig: Scintillator mirrored}
        %\caption{}
    \end{subfigure}
    %\caption{One of the two scintillator tiles used to trigger on cosmic muons passing through the active TPC volume. (a) Bare scintillator material with holes and threads allowing for mounting the tiles on the TPC as well as screwing the PCB on the scintillator. (b) The bare scintillator is wrapped with a specular reflector foil. A PCB populated with three SiPMs allows for the detection of scintillation light.}
    \caption{%One of the scintillator tiles used to trigger the TPC charge readout on cosmic muons passing through the active TPC volume.
    The bare scintillator is wrapped with a specular reflector foil and is populated with three SiPMs on a PCB.}
    \label{fig: Scintillator tiles}
\end{figure}

% SETUP: MUON TELESCOPE (SCINTILLATOR TILES)
% --------------------------------------------------------------- 
A system of two scintillator tiles with dimensions \SI{7x7x0.4}{\centi\metre}, shown in Figure~\ref{fig: Scintillator tiles}, mounted above and below the TPC provides the trigger for the charge readout.
Each tile is wrapped with a dielectric specular reflecting foil\footnote{VM2000, former name for Vikuiti ESR, 3M Inc}.
The reflector foil has a reflectance of \SI{98}[\approx]{\percent} to the visible light spectrum~\cite{Vikuiti}.
At one edge of the scintillator tile three Hamamatsu~S13360-3050VE\footnote{http://www.hamamatsu.com/us/en/product/category/3100/4004/4113/S13360-3050PE/index.html} Silicon Photo-Multipliers (SiPMs) are used to detect the scintillation light produced by crossing cosmic muons.
%A narrow printed circuit board (PCB) provides support for the SiPMs, diectly soldered to one of its sides, and terminal for the coaxial cable connectors for signal transmission soldered on the other side of the PCB.
The construction is based on that of the ArCLight light-readout system~\cite{ArCLight}.

%Cosmic muons passing through the scintillator produce scintillation light. These photons get trapped inside the cavity since the reflection coefficient of the dichroic film reaches 98\% for large incidence angles at this wavelength (THIS VALUES ARE FOR GREEN; WHICH ARE THE VALUES FOR BLUE?).
%The emission spectrum of the scintillator material matches the spectral sensitivity of the SiPMs~\cite{MPPC} closely, as shown in Figure~\ref{fig:greenbot}.

% SETUP: CRYOSTAT AND EXPECTED LAR PURITIES and HV
% ---------------------------------------------------------------
In order to perform the experiment the same cryostat and HV feed-through system as in Ref.~\cite{ElectricBreakdown_paper} were used, where a concentration of about \SI{1}{\mathrm{ppm}} of oxygen-equivalent was achieved by continuous recirculation of LAr through purification filters.
The HV power-supply\footnote{Spellmann, SL130PN150, \url{https://www.spellmanhv.com}.} readout-uncertainty is limited to \SI{0.5}{\micro\ampere}. 

\newpage
\color{white}.
\color{black}
\newpage
%%%%%%%%%%%%%%%%%%%%%%%%%%%%%%%%%%%%%%%%%%%%%%%%%%%%%%%%%%%%%
%% EXPERIMENTAL RESULTS
%%%%%%%%%%%%%%%%%%%%%%%%%%%%%%%%%%%%%%%%%%%%%%%%%%%%%%%%%%%%%
\section{Experimental results}
\label{sec: Results}
\begin{figure}[htb!]
    \centering
    \begin{subfigure}{1.0\textwidth}
        \includegraphics[draft=false,width=\textwidth]{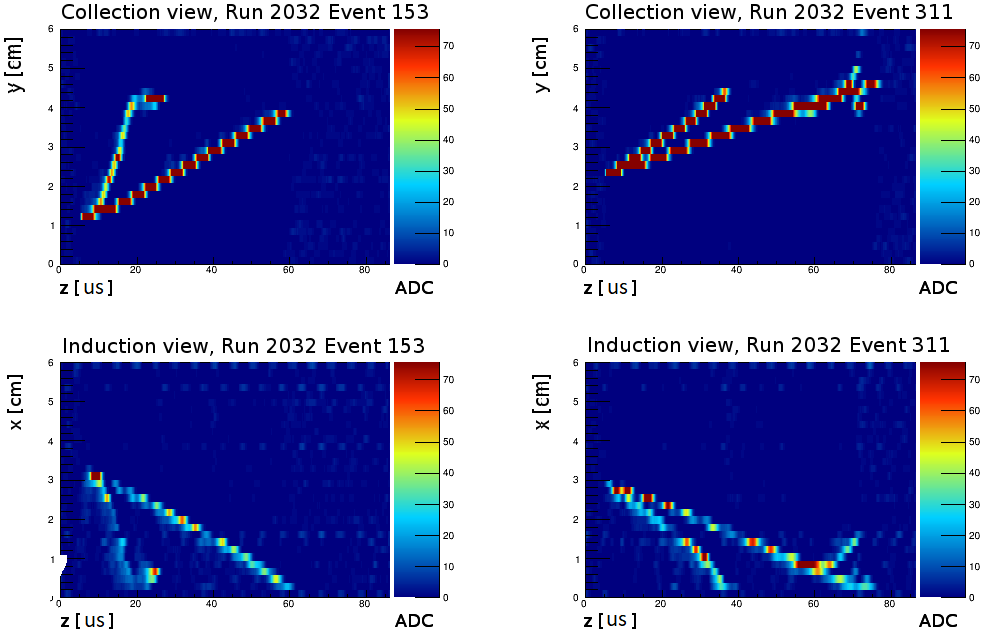}
        %\caption{caption}
        \label{fig: EventDisplay_01}
    \end{subfigure}
    ~ %add desired spacing between images, e. g. ~, \quad, \qquad, \hfill etc. 
      %(or a blank line to force the subfigure onto a new line)
%    \begin{subfigure}{0.7\textwidth}
%        \includegraphics[draft=false,width=\textwidth]{R00002033E177.png}
        %\caption{caption}
%        \label{fig: EventDisplay_02}
%    \end{subfigure}
    \caption{Event displays showing cosmic muon induced tracks within the TPC volume for an electric drift field intensity of \SI{1.5}{\kilo\volt\per\centi\metre}. $z$ denotes the drift time whereas $x$ and $y$ correspond to spatial coordinates.}
    \label{fig: Muon candidate track event display}
\end{figure}

%Only very small detector, thus: limitations.
%The design of the resistive shell is expected to be easily scalable.
The resistive shell TPC was continuously operated for $4$~days at electric field intensities up to \SI{1.5}{\kilo\volt\per\centi\metre}.
Visual inspections showed that no local LAr boiling occurred due to power dissipation across the resistive shell.
The charge-readout acquisition-trigger is based on a 2-fold coincidence between the signals coming from the top and bottom scintillator tiles.
This setup was chosen in order to acquire events originating from cosmic muons crossing the entire TPC drift length.
%Figure~\ref{fig: Muon candidate track event display} shows the collection and induction view of two cosmic induced events.
Particle tracks like those shown in Figure~\ref{fig: Muon candidate track event display} were observed across a range of various electric field intensities. 

% RESULTS: ELECTRICAL PROPERTIES OF THE RESISTIVE SHELL
% ---------------------------------------------------------------
%\subsection{Electrical properties of the resistive shell}
The electric properties of the resistive foil were measured prior to the TPC construction.
In particular, the sheet resistance of a sample with dimensions of \SI{150x15x0.05}{\milli\metre} was measured by applying a \SI{100}{\volt} bias voltage at both room temperature and in liquid nitrogen (LN2).
%To confirm that the Kapton foil has the required resistivity of about
%$\mathcal{O} \left( 1 \right) \,\mathrm{G}\Omega\mathrm{sq}^{-1}$,
%a voltage of
%\SI{100}{\volt}
%has been applied on a foil sample of
%$150~\mathrm{mm} \times 15~\mathrm{mm} \times 0.05~\mathrm{mm}$.
In LN2 a value of \SI{16}{\giga\ohm}sq$^{-1}$ was measured, significantly lower than the value of \SI{350}{\mega\ohm}sq$^{-1}$ obtained at room temperature.
This indicates a temperature dependence of the shell's sheet resistance.

\begin{figure}[htb!]
    \centering
    \begin{subfigure}{0.65\textwidth} %0.485
        \includegraphics[draft=false,width=\textwidth]{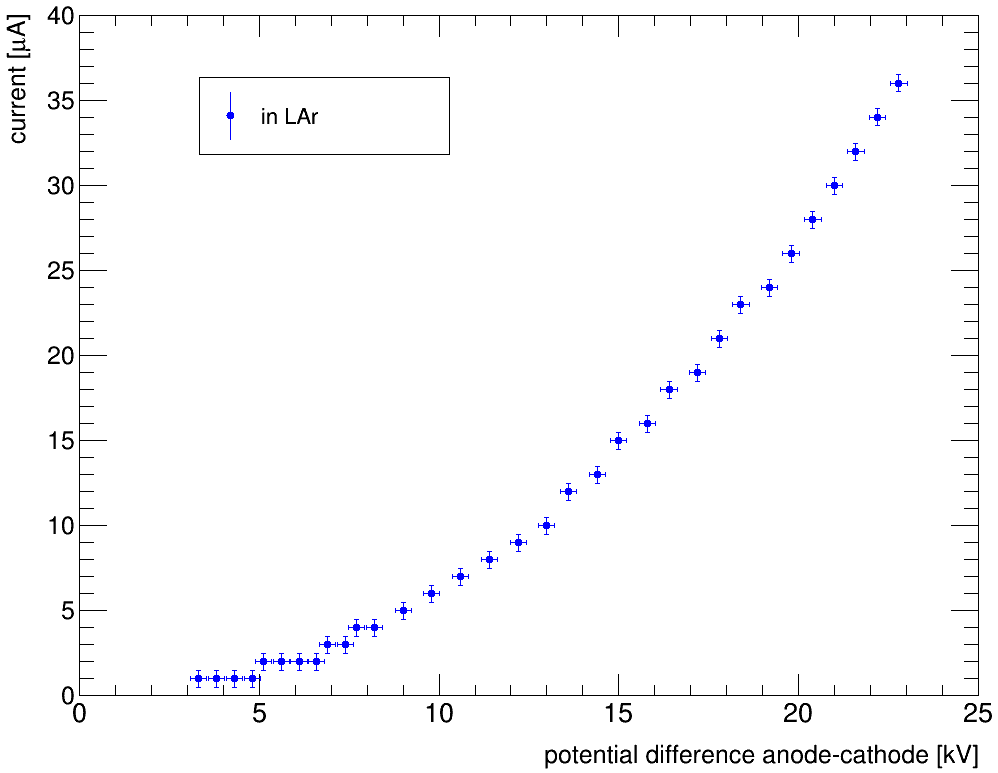}
        %\caption{df}
        \label{fig: I_V_plot}
    \end{subfigure}
    ~ %add desired spacing between images, e. g. ~, \quad, \qquad, \hfill etc. 
      %(or a blank line to force the subfigure onto a new line)
    \begin{subfigure}{0.665\textwidth} %0.485
        \includegraphics[draft=false,width=\textwidth]{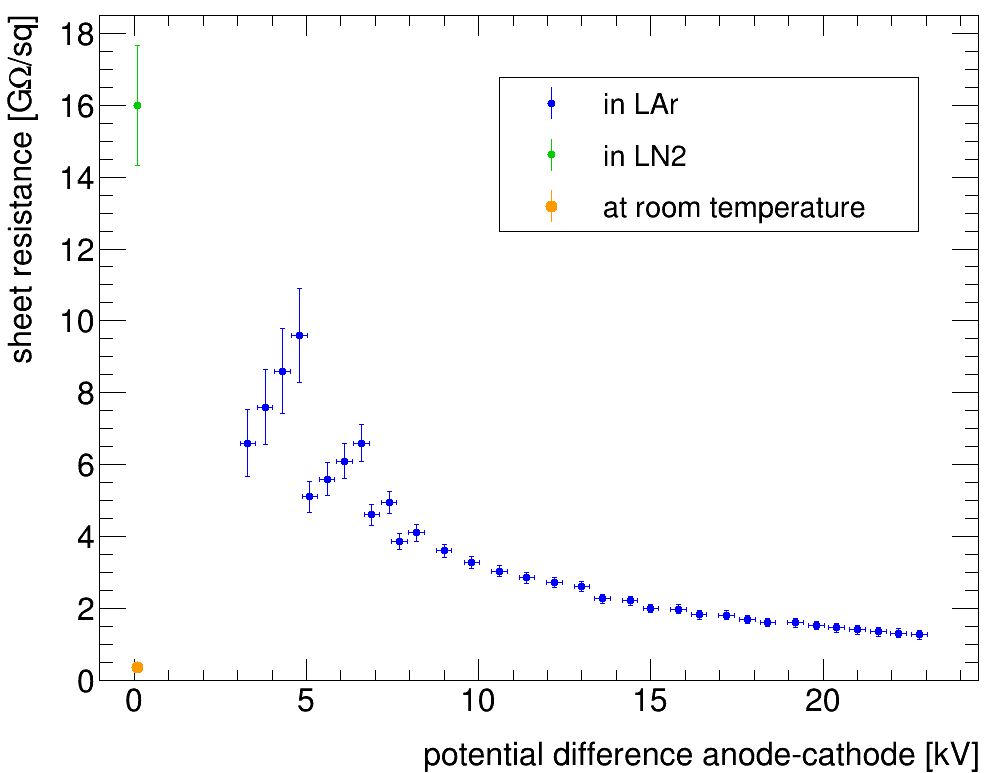}
        %\caption{df}
        \label{fig: ResistancePerSquare_V_plot}
    \end{subfigure}
    \caption{The shell's current and sheet resistance
    %measured in liquid argon, liquid nitrogen and at room temperature
    as a function of the cathode bias voltage.}
    \label{fig: Electric properties of the resistive shell}
\end{figure}

All relevant electrical properties of the resistive shell can be determined by measuring the current drawn as a function of the cathode voltage.
For this purpose, the power supply was operated up to \SI{30}{\kilo\volt} bias voltage while logging the current shown on the device.
Figure~\ref{fig: Electric properties of the resistive shell} shows the measured currents as well as the derived values for the shell's sheet resistance.
The power dissipated across the resistive shell for a field intensity of \SI{1}{\kilo\volt\per\centi\metre} was calculated to be \SI{0.23}{\watt} under the assumption of no local boiling.
%, with no local boiling observed.
At the maximum cathode bias voltage the corresponding value for the power dissipation was calculated to be \SI{0.82}{\watt}.
The sheet resistance of the resistive shell was found to be \SI{2.1}{\giga\ohm}sq$^{-1}$ for a field value of \SI{1}{\kilo\volt\per\centi\metre}, as shown in Figure~\ref{fig: Electric properties of the resistive shell}.

Due to the limited resolution of the current drawn at the HV power supply the measured sheet resistance has relatively large uncertainties for cathode bias voltages around \SI{5}{\kilo\volt} where the measured current was small.
Furthermore, to avoid singularities, values for the sheet resistance at cathode bias voltages below \SI{3}{\kilo\volt} have been excluded since the corresponding values of the current vanished.
Measuring the current with a higher accuracy would allow for a much more precise characterisation of the foil's electrical properties.

\newpage
% RESULTS: PARTICLE TRACK DISTORTIONS DUE TO NONUNIFORMITIES IN THE ELECTRIC FIELD
% ---------------------------------------------------------------
%\subsection{Particle track distortions due to non-uniformities in the electric field}
\label{sec: Results: E-field characterisation}
%The TPC charge readout system has been triggered on cosmic particles, mainly muons, passing through both scintillator tiles installed above and below the drift volume.
%For the (assay) of the detector performances cosmic muon events, featured by long and straight ionisation tracks, were selected.
%Due to their low ionisation rate (\textcolor{green}{yield?}), (\textcolor{red}{describe}) \textcolor{green}{are observed as} relatively straight tracks through the TPC active volume, as shown in figure~\ref{fig: Muon candidate track event display} [actually they are observed as straight tracks because they are MIPs and do not undergo deviation during the energy deposit...]
The performance of the resistive shell has been assessed by using cosmic muon events generating long and straight ionisation tracks.
For each event a set of charge pulses with the corresponding timing information was obtained from the waveforms acquired with all $64$ readout strips.
%of both wire planes.
%From the $64$ waveforms, coming from each of the wires of the two planes, a set of charge pulses with their drift time were obtained.
A collection of hits (3D space points) defined as the couples of unambiguous time coincident pulses from an induction ($x$~coordinate) and a collection ($y$~coordinate) strip was reconstructed for each event.
%From each of the events a collection of hits, defined as the couples of unambiguous time coincident pulses from an induction ($x$ coordinate) and a collection ($y$ coordinate) wire, was built.
Events with delta electron candidates and/or more than one track were rejected resulting in a final selection sample of $280$ muon track events with an average of $12.3$~hits per track and a most probable track length of \SI{130}{\mm}.
Since the track lengths are comparable to the liquid argon radiation length of \SI{140}{\mm}
%the width of the gaussian profile of Multiple Coulomb Scattering\cite{bib: PDG MultipleCoulombScattering and CosmicMuonEnergy}
%\begin{equation}
%\Theta_0 = \dfrac{\SI{13.6}{\mega\electronvolt}}{\beta c p} \cdot z \cdot \sqrt{\dfrac{x}{X_0}} \cdot \left[ 1 + 0.038 \cdot \ln \left( \dfrac{x}{X_0} \right) \right]
%\end{equation}
the expected hits' transverse spread
%$y_{plane}^{rms}$
due to Multiple Coulomb Scattering described in Ref.~\cite{bib: PDG MultipleCoulombScattering and CosmicMuonEnergy} is of $\mathcal{O} \left(0.1 \right) \,\mathrm{mm}$ and thus was neglected. For this calculation we assumed a muon energy of \SI{4}{\giga\electronvolt} as taken from Ref.~\cite{bib: PDG MultipleCoulombScattering and CosmicMuonEnergy}.

\begin{figure}[htb!]
\centering
\includegraphics[draft=false,width=0.5\textwidth]{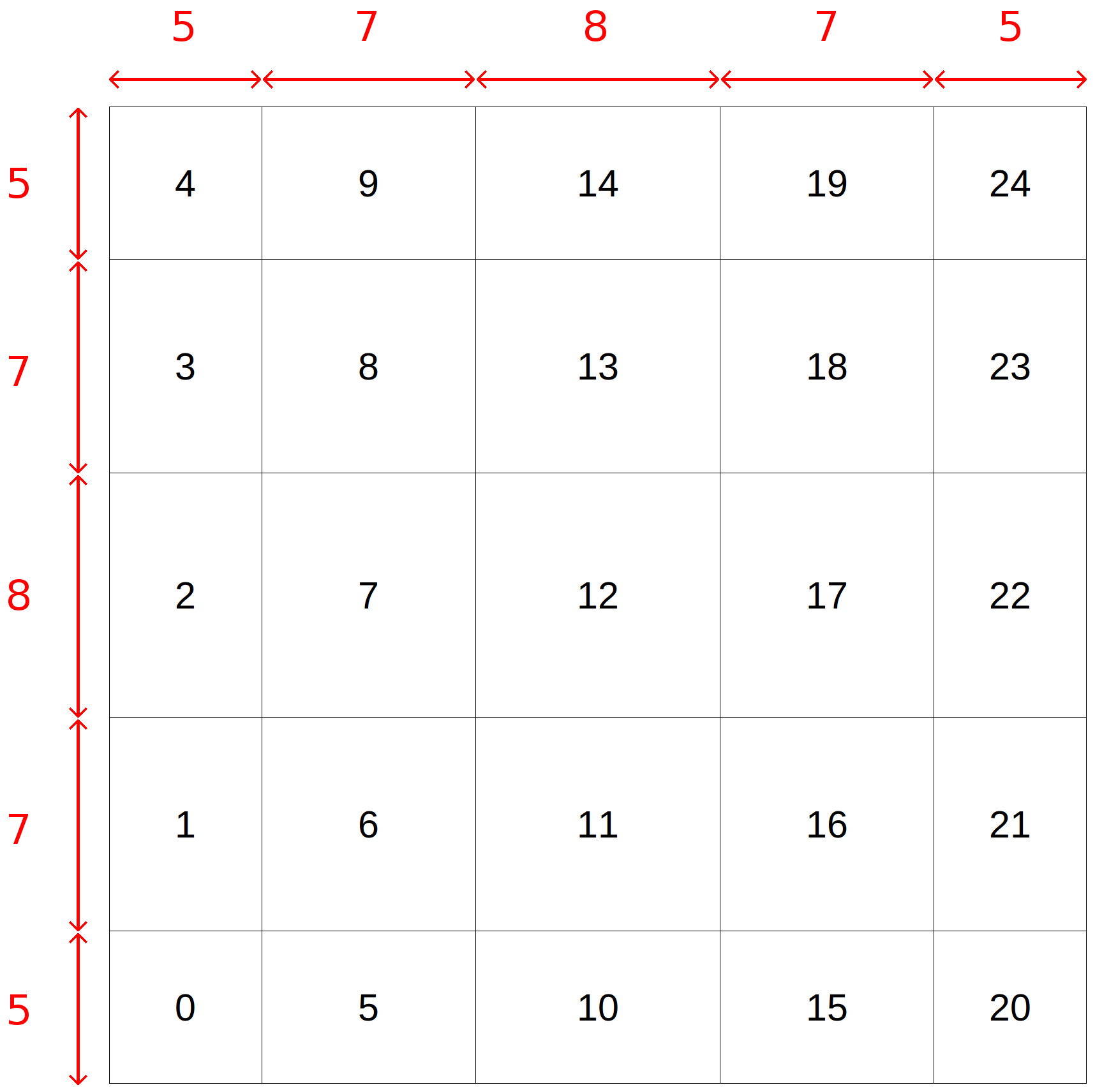}
\caption{Top view to the active volume of the TPC subdivided into $25$~regions labelled with black numbers between $0$ and $24$.
The red numbers denote the width and the length of each region in units of conducting strips.}
\label{fig: Metric and regions}
\end{figure}

Following the event selection the parameters of a straight line corresponding to the muon candidate track were determined by using the Principal Components Analysis (PCA) applied on the set of one event's hits.
%the most aligned hits, starting from the one with the shortest drift time, e.g. the hit closest to the wire readout and anode plane.
%Using a Principal Components Analysis for these hits the parameters of the straight line corresponding to the muon candidate track has been determined.
As possible electric-field non-uniformities would induce observed muon tracks with discrepancies from straight lines the hits' deviations from the obtained principal components were studied as a function of their reconstructed position.
For this study the active TPC volume was partitioned in $25$ cubic regions (cuboids between the anode and the cathode), as shown in Figure~\ref{fig: Metric and regions}.
The distributions of the three coordinates of the shortest vector (residual) from the hits to the principal component
%, which is our definition for deviation,
were obtained for each region as functions of the
%vertical coordinate
$z$-axis, corresponding to the drift direction.
Those distributions for the $8$ most central collection and induction strips are shown in Figures~\ref{fig: Residuals vs. z (2,7,12,17,22)} and \ref{fig: Residuals vs. z (10,11,12,13,14)}, respectively, as profile plots showing the mean $x$ and $y$ components of the residuals for all $280$~selected events.
%having hits in the corresponding region and in the defined $z$ bin.
Those Figures only show the statistical uncertainties.

%Some of those distributions are shown in Figure~\ref{fig: Residuals vs. z (2,7,12,17,22)} as profile plots, showing the hits' residuals accumulated over all the $280$~selected events.
%The corresponding bottom plots shows the mean residuals
%with the corresponding standard deviation (as vertical error bars)
%Also shown are the mean residuals as a function of $z$.
The $x$ and $y$ coordinates of the residuals appear to be distributed around zero with almost no dependence from the $z$ coordinate which signalises that the hits are distributed along a rather straight line.
However, the slightly bent profile plots for the residual's $x$ component in the edge regions 2 and 22 (Figure~\ref{fig: Residuals vs. z (2,7,12,17,22)}) indicate some electric-field distortions in $x$~direction near the anode ($z=0$~mm) and the cathode ($z=150$~mm).
Despite the small statistics of hits in the peripheral regions of the TPC, where a lower field uniformity is expected, the $x$ component of the residuals tend to have opposite signs for region 2 and region 22.
%, \SI{1}[\approx -]{\milli\metre} and \SI{1}[\approx +]{\milli\metre}, respectively, for hits reconstructed at a distance from the anode or cathode \SI{20}[\lesssim]{\milli\metre}.
Since for the same two regions the bespoken effect is not apparent in the profile plot of the residual's $y$ component the distortion of the electric field must be rather uniform in $y$ direction.

A similar but less pronounced behaviour is observed for the regions 7 and 17 (Figure~\ref{fig: Residuals vs. z (10,11,12,13,14)}) where the profile plots of the $y$ component of the residuals appear to be slightly bent for hits close to the electrodes.
%($z\approx0$~mm and $z\approx150$~mm).
This bending hints for small electric field distortions in $y$ direction.
In analogy to the observations made before, the residual's $x$ component for region 7 and 17 do not deviate significantly from zero, indicating vanishing field distortions in the $x$ direction.

\begin{figure}[htb!]
    \centering
    \begin{subfigure}{0.485\textwidth}
        \includegraphics[draft=false,width=\textwidth]{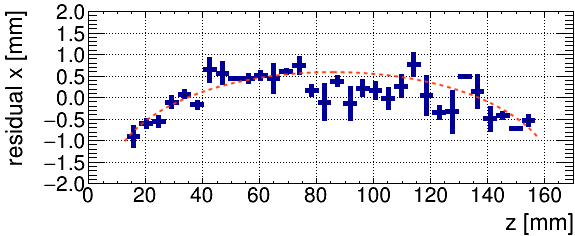}
        \caption{Region 2 (edge).}
    \end{subfigure}
    ~
    \begin{subfigure}{0.485\textwidth}
        \includegraphics[draft=false,width=\textwidth]{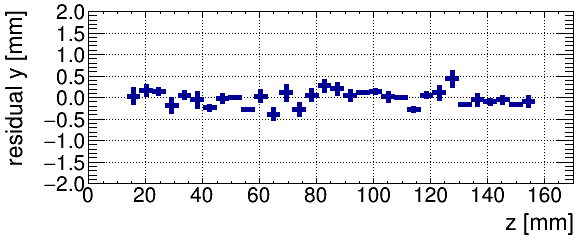}
        \caption{Region 2 (edge).}
    \end{subfigure}

    \begin{subfigure}{0.485\textwidth}
        \includegraphics[draft=false,width=\textwidth]{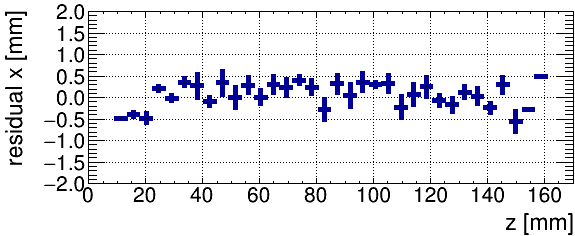}
        \caption{Region 7 (intermediate).}
    \end{subfigure}
    ~
    \begin{subfigure}{0.485\textwidth}
        \includegraphics[draft=false,width=\textwidth]{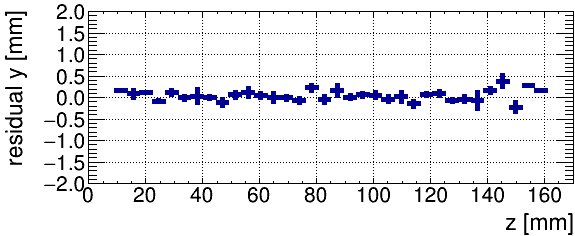}
        \caption{Region 7 (intermediate).}
    \end{subfigure}

    \begin{subfigure}{0.485\textwidth}
        \includegraphics[draft=false,width=\textwidth]{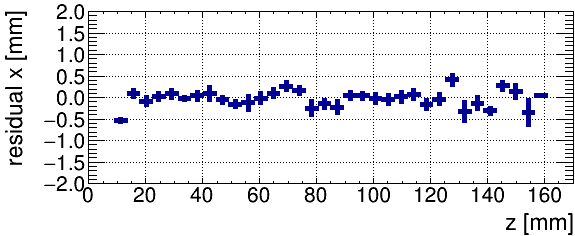}
        \caption{Region 12 (centre).}
    \end{subfigure}
    ~
    \begin{subfigure}{0.485\textwidth}
        \includegraphics[draft=false,width=\textwidth]{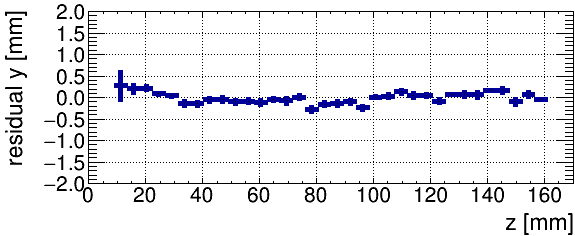}
        \caption{Region 12 (centre).}
    \end{subfigure}

    \begin{subfigure}{0.485\textwidth}
        \includegraphics[draft=false,width=\textwidth]{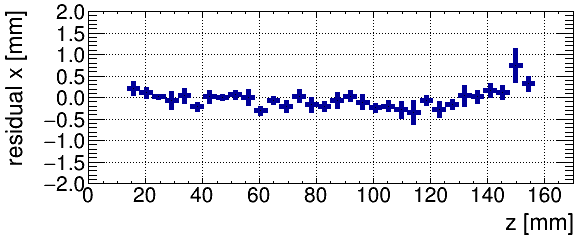}
        \caption{Region 17 (intermediate).}
    \end{subfigure}
    ~
    \begin{subfigure}{0.485\textwidth}
        \includegraphics[draft=false,width=\textwidth]{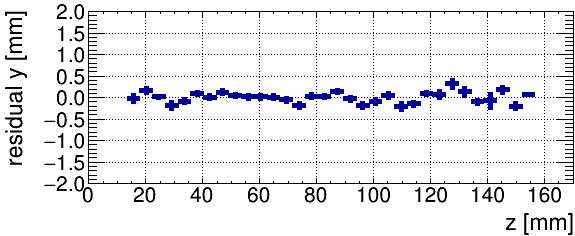}
        \caption{Region 17 (intermediate).}
    \end{subfigure}

    \begin{subfigure}{0.485\textwidth}
        \includegraphics[draft=false,width=\textwidth]{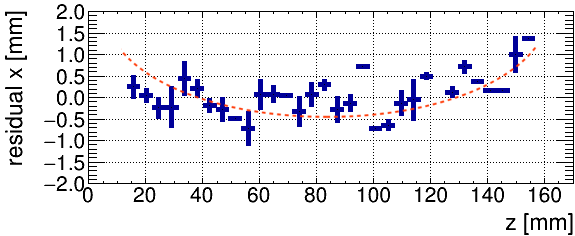}
        \caption{Region 22 (edge).}
    \end{subfigure}
    ~
    \begin{subfigure}{0.485\textwidth}
        \includegraphics[draft=false,width=\textwidth]{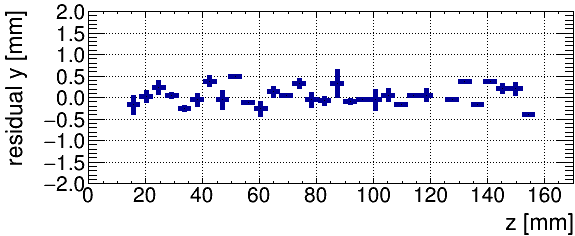}
        \caption{Region 22 (edge).}
    \end{subfigure}
    \caption{The residual's $x$ and $y$ components as a function of the drift distance $z$ for five adjacent regions. Dashed orange lines are used to illustrate the tendency.}
    \label{fig: Residuals vs. z (2,7,12,17,22)}
\end{figure}

\begin{figure}[htb!]
    \centering
    \begin{subfigure}{0.485\textwidth}
        \includegraphics[draft=false,width=\textwidth]{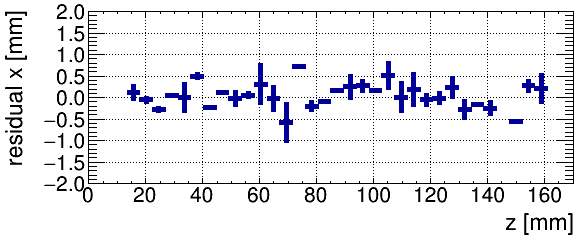}
        \caption{Region 10 (edge).}
    \end{subfigure}
    ~
    \begin{subfigure}{0.485\textwidth}
        \includegraphics[draft=false,width=\textwidth]{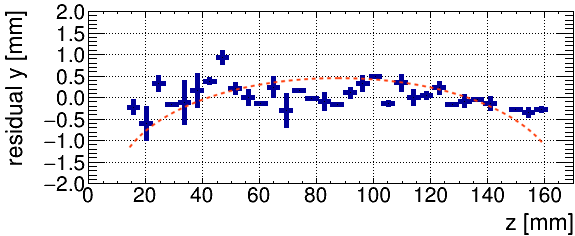}
        \caption{Region 10 (edge).}
    \end{subfigure}

    \begin{subfigure}{0.485\textwidth}
        \includegraphics[draft=false,width=\textwidth]{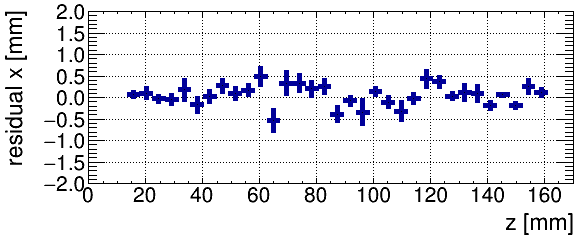}
        \caption{Region 11 (intermediate).}
    \end{subfigure}
    ~
    \begin{subfigure}{0.485\textwidth}
        \includegraphics[draft=false,width=\textwidth]{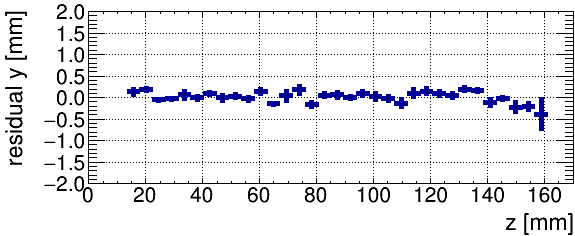}
        \caption{Region 11 (intermediate).}
    \end{subfigure}

    \begin{subfigure}{0.485\textwidth}
        \includegraphics[draft=false,width=\textwidth]{residuals_x_vs_z_12_2.png}
        \caption{Region 12 (centre).}
    \end{subfigure}
    ~
    \begin{subfigure}{0.485\textwidth}
        \includegraphics[draft=false,width=\textwidth]{residuals_y_vs_z_12_2.png}
        \caption{Region 12 (centre).}
    \end{subfigure}

    \begin{subfigure}{0.485\textwidth}
        \includegraphics[draft=false,width=\textwidth]{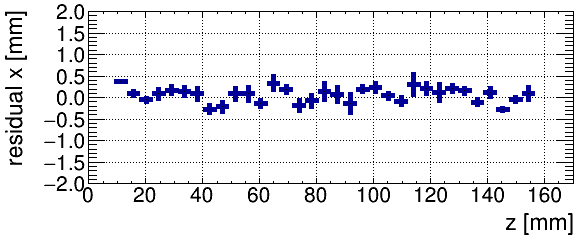}
        \caption{Region 13 (intermediate).}
    \end{subfigure}
    ~
    \begin{subfigure}{0.485\textwidth}
        \includegraphics[draft=false,width=\textwidth]{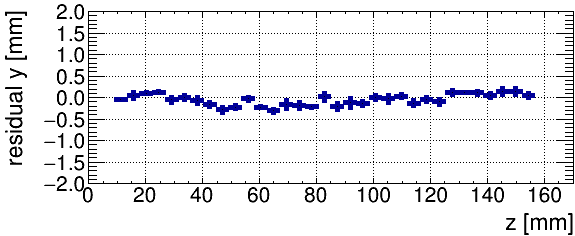}
        \caption{Region 13 (intermediate).}
    \end{subfigure}

    \begin{subfigure}{0.485\textwidth}
        \includegraphics[draft=false,width=\textwidth]{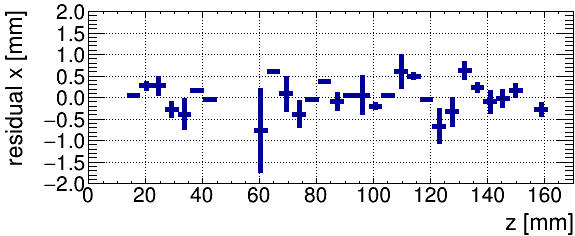}
        \caption{Region 14 (edge).}
    \end{subfigure}
    ~
    \begin{subfigure}{0.485\textwidth}
        \includegraphics[draft=false,width=\textwidth]{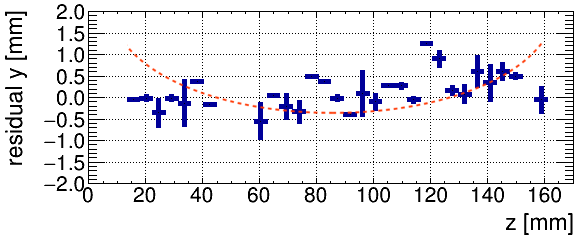}
        \caption{Region 14 (edge).}
    \end{subfigure}
    \caption{The residual's $x$ and $y$ components as a function of the drift distance $z$ for five selected regions. Dashed orange lines are used to illustrate a possible tendency.}
    \label{fig: Residuals vs. z (10,11,12,13,14)}
\end{figure}

\newpage
\color{white}.
\color{black}
\newpage
\color{white}.
\color{black}
\newpage

\begin{figure}[htb!]
\centering
\includegraphics[draft=false,width=0.9\textwidth]{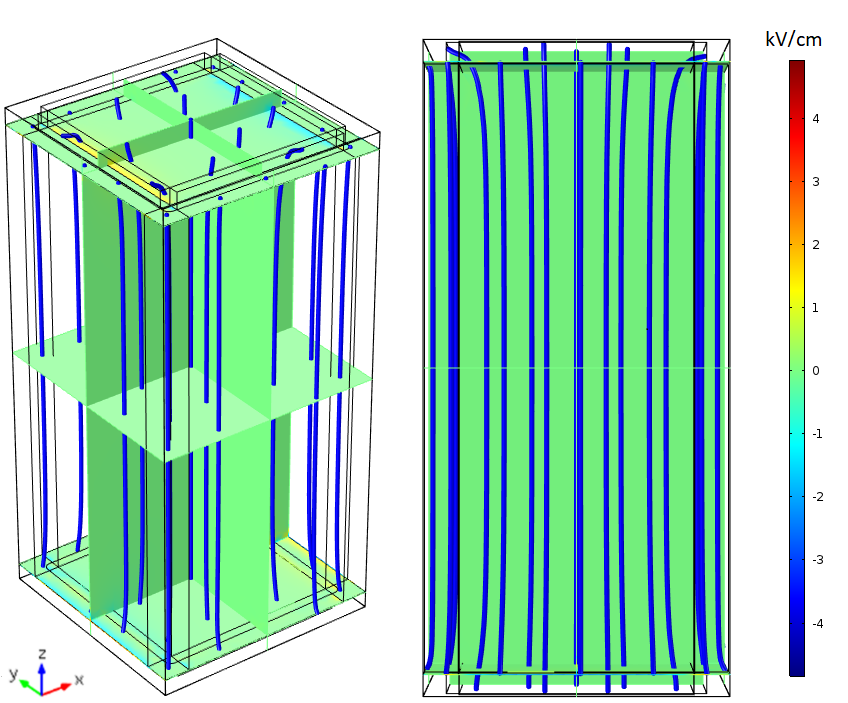}
\caption{Electric-field lines (blue tubes) and the electric-field distortions in $x$ direction only (slices), simulated with COMSOL Multiphysics. The distortions in $x$ direction are largest, $\mathcal{O} \left(100 \right) \,\mathrm{kV}$, very close to the iron frames (yellow and light blue colors) but vanish for most of the active volume (indicated by the green slices).}
\label{fig: COMSOL simulated electric field}
\end{figure}

The observation that some hits near the anode or the cathode plane have larger deviations from the principal component than hits far from the electrodes motivated us to simulate the electric field within the drift volume of the TPC.
For this purpose the COMSOL Multiphysics\footnote{\url{https://www.comsol.com}} code version 5.2 was used.
As a result, the electric-field lines as well as its $x$~components within the TPC drift volume are shown in Figure~\ref{fig: COMSOL simulated electric field}.
It appears that the magnitude of the electric field is a few percent higher at the edge of the drift volume than in the middle.
The field lines therefore are bent towards the centre of the TPC's active volume.
Furthermore, the field's $x$ and $y$ components are largest near the anode and the cathode and rapidly vanish at larger distances from the electrodes.
The maximum electric-field distortions in $x$ (similar for the $y$ direction) are about $\mathcal{O}\left(1\right) \,\mathrm{kV/cm}$ at the very top and bottom of the drift volume.

When simulating the TPC without any iron frame the electric-field lines are no longer bent but describe straight lines between the anode and the cathode.
%The production of the electric field distortions therefore are ascribed to the two iron frames.
Hence, the observed distortions can likely be ascribed to the two iron frames and not to some features of the resistive shell.

As the ionisation electrons drift along the electric field lines, electrons produced at the edges of the TPC drift volume tend to move more outwards and are expected to have larger deviations from the principal component than electrons produced at the centre of the TPC.
In addition, electrons that start to drift from regions closer to the electrodes will be reconstructed with smaller $x$-$y$ deviation than those starting from the middle of the TPC.
This is qualitatively consistent with the bent profile plots of the $x$ and $y$ components of the residuals observed in the most peripheral regions 2, 22, 10 and 14 (Figures~\ref{fig: Residuals vs. z (2,7,12,17,22)} and \ref{fig: Residuals vs. z (10,11,12,13,14)}).
However, for a deeper quantitative comparison between the predictions from simulations and the observed deviations a higher statistics of hits in the outer regions of the active volume, where these effects are stronger, would be needed.
\section{Conclusion}
\label{sec: Conclusion}
A prototype liquid argon TPC instrumented with the novel approach of the resistive shell for field shaping was built and tested at LHEP, University of Bern.
%A novel design for the field shaping in liquefied noble gas TPCs, the resistive shell, is proposed and a prototype TPC using this technology has been built.
The shell consists of a highly resistive, $\mathcal{O}\left(10^9\right) \,\mathrm{\Omega/sq}$, carbon-loaded Kapton foil with a thickness of \SI{50}{\micro\metre}.
%Experiments in LAr showed that the cathode as well as the resistive shell can be built with the discussed foil to produce and shape an electric field within the shell.
The measured electrical properties and performance of the TPC prototype showed that this innovative design is suitable for applications that require low-power consumption, minimisation of the TPC material for field shaping and maximisation of the active TPC volume.
%Since the resistive shell is best suited to maximize the active TPC volume and to minimize inactive dense material in a TPC, this new technology is a promising alternative to traditional field shaping structures.
Furthermore, the high resistivity of the shell and the reduced number of components
%, in the case of an electric breakdown such a design
reduces the power release
%in the discharge
as well as the potential number of failure points during operating the TPC at high bias voltages.
%Furthermore, the shell is expected to limit the power dissipation in the case of an electric breakdown and, with the reduced number of components, reduces the risk of possible failures during the operation of a TPC.
Detailed studies for the characterisation of the resistive foil's power dissipation in the case of an electric breakdown are still missing and will be object of future experimental works employing similar modalities as in Ref.~\cite{ElectricBreakdown_paper}.
Finally, compared to a conventional field cage with a resistor chain the resistive shell minimizes the heat load per surface area and thus reduces local boiling of the noble liquid.
%Finally, the resistive shell would simplify the feasibility of a modular design of large noble liquid TPCs such as proposed by the ArgonCube collaboration.

%ALSO MENTION: FOR NEXT EXPERIMENT IT WOULD BE NICE TO HAVE A SCINTILLATOR WITH SOME SPATIAL RESOLUTION. THEN, THE TRACK RECONSTRUCTION CAN BE CONSTRAINED AND IMPROVED.

%The new resistive shell technology has several advantages:
%\begin{itemize}
%    \item The active volume of a TPC can be maximized when minimising dense and inactive material within the TPC.
%    \item Due to a limited power dissipation in the case of electric brakdowns the risks for the TPC operation are reduced.
%    \item The reduced number of components minimises the number of possible points of failure during the TPC operation.
%    \item Compared to a traditional field cage with a resistor chain the resistive shell decreases the heat load per surface area and thus reduces local boiling.
%\end{itemize}
%Furthermore, the resistive shell would simplify the feasibility of a modular design of large liquid noble gas TPCs such as the ArgonCube collaboration (REF.) proposes.

During a measurement campaign of about \SI{5}{\mathrm{days}} in July 2018, straight ionisation tracks induced by cosmic muons were detected.
Since the prototype TPC is rather small, with a drift-field volume of only \SI{15x7x7}{\cm}, the characterisation of the electric-field uniformity obtained with the experiment is limited.
Further limitations come from the small sample of only $280$~analysed muon candidate tracks.
However, the analysis discussed in Section~\ref{sec: Results} shows mean deviations from the hits to the track's principal component of less than \SI{1}{\milli\metre}, which are largest in regions close to the edges of the TPC and near the anode or the cathode plane.
This can be qualitatively explained with the COMSOL simulated electric-field distortions introduced by two iron frames connecting the anode and the cathode with the resistive shell.
According to simulations the electric-field distortions vanish when removing the frames showing that this effect does not depend on any specific features of the adopted technique.
%, hence the uniformity of the electric field could be significantly better if no iron frames were used.
%However, an analysis to quantitatively characterise the field distortions and those induced on the reconstructed tracks is beyond the purposes of this work.

%An analysis to quantitatively characterize the electric field distortions using muon candidate tracks is ongoing. This analysis uses the COMSOL simulated electric field lines to propagate every hit from the charge readout plane back to the space point where the ionisation must have happened.
%This approach aims for correcting for the the electric field distortions. After those correction have been applied, a similar study as presented in Section~\ref{sec: Results: E-field characterisation} can be performed to characterize nonuniformities in the electric field.

In order to further characterise the capabilities of the resistive shell to shape electric fields in TPCs a larger device without iron frames and able to collect larger samples of straight muon tracks will be required.
Another possibility to characterise the electric-field distortions would be the use of a steerable UV laser, such as presented in Ref.~\cite{Steerable_UV_laser_paper, Laser_drift_field_characterisation}.

We observed that the sheet resistance of the shell depends in a non-linear way on the applied bias voltage.
Since this behaviour would affect the heat load and thus the diffusion of the noble liquid when the applied voltage is changed, future experiments should test and quantify the long term stability of the foil's electric properties when a constant bias voltage is applied to it.

The tested TPC design can be easily scaled to several square meters of surface area of the shell as needed by the next generation of detectors for neutrino physics as well as for neutrino-less double beta decay experiments and direct dark matter searches.
%$0\nu\beta\beta$ and direct dark matter searches.
For this purpose we will adopt this novel TPC technology for the upcoming ArgonCube 2x2 Demonstrator
%~\cite{ArgonCube_LOI}
as well as for the possible development of one of the far-detector modules of the DUNE experiment.
\section*{Acknowledgements}
Many thanks to the workshop of LHEP for the engineering and technical support, Jan Christen, Roger Hänni, Lorenzo Meier, and Camilla Tognina.
Furthermore, we acknowledge financial support from the Swiss National Science Foundation and from the Canton of Bern, Switzerland.
%We acknowledge financial support from the Swiss National Science Foundation.
%The Bern group is supported by the Swiss National Science Foundation
%grant (NUMBER???)
%and the Canton of Bern, Switzerland.
%The Fermilab group is supported by the U.S.~Department of Energy under contract NUMBER.
%Do not mention any grant related with Fermilab (see email from Ting in Jan. 10th 2019)

\end{document}